\documentclass[10pt,a4paper,twocolumn,english,amssymb,nofootinbib,superscriptaddress,showpacs]{revtex4}
\usepackage{lmodern}

\usepackage[T1]{fontenc}
\usepackage[latin9]{inputenc}
\usepackage{color}
\usepackage{babel}
\usepackage{amsmath}
\usepackage{amssymb}
\usepackage{esint}
\usepackage[unicode=true,pdfusetitle,
 bookmarks=true,bookmarksnumbered=false,bookmarksopen=false,
 breaklinks=false,pdfborder={0 0 1},backref=false,colorlinks=false]
 {hyperref}

\makeatletter

\pdfpageheight\paperheight
\pdfpagewidth\paperwidth

\@ifundefined{textcolor}{}
{%
 \definecolor{BLACK}{gray}{0}
 \definecolor{WHITE}{gray}{1}
 \definecolor{RED}{rgb}{1,0,0}
 \definecolor{GREEN}{rgb}{0,1,0}
 \definecolor{BLUE}{rgb}{0,0,1}
 \definecolor{CYAN}{cmyk}{1,0,0,0}
 \definecolor{MAGENTA}{cmyk}{0,1,0,0}
 \definecolor{YELLOW}{cmyk}{0,0,1,0}
}

\usepackage{babel}
\@ifundefined{definecolor}{\usepackage{color}}{}
\@ifundefined{definecolor}{\usepackage{color}}{}
\usepackage{babel}

\@ifundefined{definecolor}{\usepackage{color}}{}
\usepackage{babel}

\pdfpageheight\paperheight
\pdfpagewidth\paperwidth

\@ifundefined{textcolor}{}{%
 \definecolor{BLACK}{gray}{0}
 \definecolor{WHITE}{gray}{1}
 \definecolor{RED}{rgb}{1,0,0}
 \definecolor{GREEN}{rgb}{0,1,0}
 \definecolor{BLUE}{rgb}{0,0,1}
 \definecolor{CYAN}{cmyk}{1,0,0,0}
 \definecolor{MAGENTA}{cmyk}{0,1,0,0}
 \definecolor{YELLOW}{cmyk}{0,0,1,0}
 }

\@ifundefined{definecolor}{\usepackage{color}}{}
\usepackage{latexsym}\usepackage{bm}

\makeatother

\begin{document}

\title{Event Horizon Detecting Invariants }

\author{Aydin Tavlayan}

\email{aydint@metu.edu.tr}

\selectlanguage{english}%

\affiliation{Department of Physics,\\
 Middle East Technical University, 06800 Ankara, Turkey}
\author{Bayram Tekin}

\email{btekin@metu.edu.tr}

\selectlanguage{english}%

\begin{abstract}
\noindent Some judiciously chosen local curvature scalars can be used to invariantly characterize 
event horizons of black holes in $D > 3$ dimensions, but they fail for the three dimensional    
Ba\~nados-Teitelboim-Zanelli (BTZ) black hole since all curvature invariants are constant. Here we provide an invariant characterization of the event horizon of the BTZ black hole using the 
curvature invariants of codimension one hypersurfaces instead of the full spacetime. Our method is also applicable to black holes in generic dimensions but is most efficient in three, four, and five dimensions. We give four dimensional Kerr, five dimensional Myers-Perry and three dimensional warped-anti-de Sitter, and the three dimensional asymptotically flat black holes as examples.    
\end{abstract}
\maketitle

\section{Introduction}
The event horizon of a black hole is a codimension one null hypersurface, which constitutes the boundary of the black hole region from which casual geodesics cannot reach future null infinity \cite{Eric}. As such, the event horizon is highly non-local, and so one a priori needs the full knowledge of the spacetime to locate it. In this respect, for example, the image obtained by the Event Horizon Telescope (EHT) contains the environment of the black hole as well as the codimension two cross-section of the event horizon but not the event horizon which is a 2+1 dimensional hypersurface. 

Given a metric representing a spacetime,  the usual search for an event horizon proceeds as follows: one assumes that the event horizon ${\mathcal H}$ is a smooth level set of a single function $F$ satisfying $ g^{\mu \nu}  \partial_\mu  F \partial_\nu F =0$ which clearly requires
well-chosen coordinates in spacetime. Instead of this coordinate-dependent method, there have been some recent developments initiated by the work of Abdelqader and Lake \cite{Lake} who gave an invariant characterization of the Kerr black hole, significantly extending the earlier method of Karlhede et al. \cite{Karlhede} which works for the case of the Schwarzschild black hole.  In \cite{Lake} the following set of curvature scalars was suggested as a basis to detect the location of the event horizon and the ergo-surface as well as to define some other properties, such as the mass and the spin of the black hole
\begin{eqnarray}
 &  &I_1=C_{\mu\nu\alpha\beta}C^{\mu\nu\alpha\beta}, \hskip 1.2 cm I_2= {}^\ast C_{\mu\nu\alpha\beta}C^{\mu\nu\alpha\beta},    \nonumber\\
 &  &I_3=\nabla_{\rho}C_{\mu\nu\alpha\beta}\nabla^{\rho}C^{\mu\nu\alpha\beta},    \hskip 0.3 cm
 I_4=\nabla_{\rho}C_{\mu\nu\alpha\beta}\nabla^{\rho} {}^\ast C^{\mu\nu\alpha\beta},   \nonumber \\
 &  &I_5=k_{\mu}k^{\mu}, \,\,\,\,\, 
 I_6=l_{\mu}l^{\mu},  \,\,\, \,\,\,\,\,\,\,
 I_{7}=k_{\mu}l^{\mu}
\label{ilk}
\end{eqnarray}\\
where $ C_{\mu\nu\alpha\beta} $ is the Weyl Tensor, $ {}^\ast C_{\mu\nu\alpha\beta} $ is its left dual and the two covectors defining the last three scalars are given as $ k_{\mu}=-\nabla_{\mu}I_1 $, and $ l_{\mu}=-\nabla_{\mu}I_2 $. To detect the location of the event horizon of the Schwarzschild black hole $I_3$ is sufficient: Namely, that scalar is positive outside the event horizon, vanishes on the event horizon, and is negative inside the horizon. For the Kerr black hole, none of the above invariants are enough to locate the event horizon. Hence in  \cite{Lake}, the following nonlinear combination was found 
\begin{equation}
Q_{2}=\dfrac{1}{27}\frac{I_{5}I_{6}-I_{7}^{2}}{(I_{1}^{2}+I_{2}^{2})^{5/2}} 
\end{equation}\\
which vanishes on the event horizon of the Kerr black hole. In constructing this invariant, three syzygies that are valid for the Kerr black hole were found among the seven invariants (\ref{ilk}), reducing the independent invariants to four, which are sufficient to invariantly describe the mass, spin, and the $r,\theta$ coordinates of the cohomogeneity of the metric.  

As can be deduced from the above discussion, for each kind of black hole, one must judiciously define a curvature scalar that can be used to detect the event horizon. This ad hoc  state of  affairs was remedied  by Page and Shoom \cite{Page}, who proposed a systematic way of constructing such invariants.  The crux of their idea boils down to the fact that the event horizon is a null hyperspace  with a degenerate metric; as such, the determinant of the hypersurface metric vanishes on the event horizon. However, this statement can be invariantly stated not with the coordinates (which are scalar functions defined on open subsets of the spacetime manifold) but with some other scalar functions built from curvature invariants. The execution of their procedure is as follows: the squared norm of the wedge product of $n$ gradients of independent local smooth curvature invariants would vanish on the horizon of any stationary black hole, where $n$ is the local cohomogeneity of the spacetime.

The Page-Shoom method is powerful, but it does not work for the critical example of the 2+1 dimensional BTZ black hole \cite{BTZ}, which is a solution of Einstein's theory with a negative cosmological constant, as well as any other three dimensional gravity theory which admits  $AdS_3$ as a solution. The reason for this failure is simple to understand: in three dimensions, the Riemann tensor is algebraically related to the Einstein tensor as $R_{\mu \alpha \nu \beta} = \epsilon_{\mu \alpha \sigma}  \epsilon_{\nu \beta \rho} G^{\sigma \rho}$: hence whenever the Einstein tensor vanishes, the Riemann tensor vanishes along with all the curvature invariants; and whenever the Einstein tensor is proportional to the metric tensor (as in the case of the BTZ black hole) the Riemann tensor is locally maximally symmetric as in $AdS_3$ and all the curvature invariants are just constants. The conclusion is that the spacetime curvature scalars cannot be used to detect the event horizon of the BTZ black hole. This prompted us to search for a new approach that we describe in this work. 

\section{Curvature invariants of the hypersurface}

Our method, which works for the BTZ black hole and other black holes in three, four, and five dimensions, 
is a variant of the method of Page and Shoom \cite{Page}, but only uses one curvature invariant.    Beyond five dimensions,  our method will still be valid, but one needs more than one curvature invariant as such, it is not more advantageous than the method of \cite{Page}, which also generically requires more than one curvature invariant.

The main idea of our proposal is as follows: as was realized in \cite{Page} in a spacetime with a stationary black hole, there is always a symmetry that can be used to split the tangent space of the spacetime into two parts. Let $G$ be the local symmetry group with $m$ dimensional maximal orbits. Hence the local cohomogeneity has $n = D-m$ dimensions. Considering a set of gradients $\{  d S^{i}\}$ built from $n$ functionally independent {non-constant } curvature scalars, their wedge product, say $W$, (which is proportional to the volume form on the cohomogeneity space) will vanish on the event horizon since the Hodge dual of $W$ vanishes there. This can be used to detect the event horizon. However, clearly, for the case of the BTZ black hole, there is no such non-constant curvature scalar, as shown above. One possible way out of this problem is to reduce the local cohomogeneity. Especially, if we can reduce the local cohomogeneity to one, we can use just one curvature invariant, which vanishes on the horizon. In order to reduce the local cohomogeneity, we embed the $D-1$ dimensional surface $\Sigma$ to the  $D$-dimensional spacetime. It is essential to understand that this $\Sigma$ is not the event horizon ${\mathcal H}$. Let us carry out this procedure for the BTZ black hole first and then to other black holes. 

\section{BTZ BLACK HOLE}

In ($ t,r,\phi $) coordinates, the BTZ metric is given as
\begin{eqnarray}
& & g_{\mu\nu}=\left(
\begin{array}{ccc}
m-\frac{r^2}{\ell^2} & 0 & -\frac{j}{2} \\
0 & \frac{1}{\frac{j^2}{4 r^2}+\frac{r^2}{\ell^2}-m} & 0 \\
-\frac{j}{2} & 0 & r^2 \\
\end{array}
\right)
\label{fullbtz}
\end{eqnarray}\\
where $ m $  and  $ j$ are mass and angular momentum, respectively; and $\ell$ is the $AdS_3$ radius.   In this coordinate system, the coordinates $ t $, and $ \phi $ have associated Killing symmetries, $\partial_t$ and $\partial_\phi$ respectively. The event horizon can be easily found as the largest root of $g^{rr}=0$ or equivalently as the location at which the time-like Killing vector $\xi= \partial_t + \Omega \partial_\phi$ becomes null  where  $\Omega = -g_{t\phi}/g_{\phi\phi}$ which becomes $\Omega = j/(2 r_+^2)$ at the event horizon. Both results yield 
\begin{equation}
r_+^2 = \frac{ m \ell^2}{2} \left( 1+ \sqrt{1 - \frac{ j^2}{ \ell^2 m^2}} \right),
\label{root1}
\end{equation}
with the positive root giving the event horizon.

Let us now find this in a coordinate independent way. The cohomogeneity of this spacetime is one. Keeping this intact, we need to choose a hypersurface which has at least one non-constant curvature invariant. For this purpose we can choose the $t = t_0$ or 
$ \phi=\phi_{0}$ to define the hypersurface (note again that it is not the event horizon itself). Both options are valid and yield the same result, let us choose the latter, then the induced metric is 
\begin{eqnarray}
& & \gamma_{i j}=\left(
\begin{array}{ccc}
m-\frac{r^2}{\ell^2} & 0  \\
0 & \frac{1}{\frac{j^2}{4 r^2}+\frac{r^2}{\ell^2}-m}  \\
\end{array}
\right)
\label{btzsurface}
\end{eqnarray}
in the induced coordinates $ (t,r). $ This metric has a symmetry only in the $t$ coordinate. So, the cohomogeneity is still one, as expected. The Kretschmann invariant computed for the induced  metric (\ref{btzsurface}) becomes, 
\begin{equation}
{}^{\Sigma}I_1={}^{\Sigma}R_{i j k l}{}^{\Sigma}R^{i j k l}=\frac{\left(j^2 \ell^2-4 \left(r^2-\ell^2 m\right)^2\right)^2}{4 \left(\ell^3 m-\ell r^2\right)^4},
\end{equation} 
from which one can calculate the horizon detecting invariant 
\begin{equation}
{}^{\Sigma}I_{5}=\nabla_{m}{}^{\Sigma}I_1 \nabla^{m} {}^{\Sigma}I_1
\label{gerekli}
\end{equation}\\
which reads explicitly as
\begin{equation}
{}^{\Sigma}I_{5}=\frac{j^4 \left(j^2 \ell^2-4 \ell^2 m r^2+4 r^4\right) \left(j^2 \ell^2-4 \left(r^2-\ell^2 m\right)^2\right)^2}{\ell^6 \left(r^2-\ell^2 m\right)^{10}}.
\end{equation}\\
whose largest real root, ${}^{\Sigma}I_{5}(r=r_+)=0$,  is exactly the event horizon given as \ref{root1}.

In fact for the BTZ black hole, the hypersurface version of the curvature invariant given in\cite{Karlhede} can also be used to detect the event horizon as a somewhat simpler way. One can show that for the hypersurface metric (\ref{btzsurface}) one has 
\begin{eqnarray}
{}^{\Sigma}I_{3}=&&\nabla_{m}{}^{\Sigma}R_{i j k l}\nabla^{m}{}^{\Sigma}R^{i j k l} \label{gerekli2} \\
=&&\frac{j^4 \left(j^2 \ell^2-4 \ell^2 m r^2+4 r^4\right)}{\ell^2 \left(r^2-\ell^2 m\right)^6}\nonumber
\end{eqnarray}
which vanishes only on the event horizon and on the inner horizon. Therefore, it is a very useful  tool for event horizon detection.

\section{Warped $AdS_3$ black hole}

The method described above is geometric, in the sense that  it relies on the metric and not on the underlying field equations. Therefore, it can be applied to other metrics that solve field equations that are different from general relativity. Let us consider the case of the warped $AdS_3$ black hole in the ($ t,r,\phi $) coordinates \cite{Nutku,Gurses,Li}.
\begin{eqnarray}
ds^2=&&-\ell^{2}dt^{2}+\frac{\ell^2}{\left(\nu^2+3\right) (r-p) (r-q)}dr^{2}+ \nonumber \\ 
&& +\frac{1}{4} \ell^2 r f d\phi^{2}  \\ 
&& -\ell^2 \left(\sqrt{p q \left(\nu^2+3\right)}-2 r \nu\right)dtd\phi. \nonumber
\end{eqnarray}\\
where the function $f$ is given as 
\begin{equation}
f:= \left(\nu^2+3\right) (p+q)-4 \nu \sqrt{p q \left(\nu^2+3\right)}+3 r \left(\nu^2-1\right).\nonumber
\end{equation}
The event horizon  is at $ r_{+}=p, $ and the inner horizon is  at $r_{-}=q$.
This metric solves the topologically massive gravity \cite{DJT} for  $ \nu=\mu\dfrac{\ell}{3}$ 
\begin{equation}
R_{\mu \nu} - \frac{1}{2} R g_{\mu \nu} - \frac{1}{\ell^2} + \frac{1}{ \mu} C_{\mu \nu}=0 ,
\end{equation}
where $C_{\mu \nu}$ is the Cotton tensor and $\mu$ is the topological mass. Even though the Ricci tensor and the Riemann tensor of this spacetime are different forms of the maximally symmetric $AdS_3$, the curvature invariants of the full spacetime cannot detect the event horizon because all the curvature scalars are constant. In fact, explicit forms of some of the curvature invariants are  given as 
\begin{eqnarray}
&&R_{\mu \nu \rho \sigma} R^{\mu \nu \rho \sigma }= \frac{12 \left(2 \nu^4-4 \nu^2+3\right)}{\ell^4} \nonumber \\
&&R_{\mu \nu} R^{\mu \nu}=\frac{6 \left(\nu^4-2 \nu^2+3\right)}{\ell^4}\nonumber \\
&&\nabla_{\sigma}R_{\mu \nu} \nabla^{\sigma}R^{\mu \nu}=-\frac{36 \nu^2 \left(\nu^2-1\right)^2}{\ell^6}\nonumber \\
&&\nabla_{\sigma}R_{\mu \nu\rho\gamma} \nabla^{\sigma}R^{\mu \nu \rho\gamma}=-\frac{144 \nu^2 \left(\nu^2-1\right)^2}{\ell^6}
\end{eqnarray}
which are not useful to detect the event horizon.
Therefore we resort to our method, but there is a subtle issue here. As in the case of the BTZ black hole, there are two Killing symmetries for the warped $AdS_3$ black hole and two possible ways of choosing the hypersurface by keeping the cohomogeneity constant. However, only one of these two ways yields a hypersurface with non-constant curvature  invariants. In fact, if we define the hypersurface as $ \phi=\phi_{0}$, we obtain a flat metric, that is not useful at all. The other option, that is choosing the hypersurface defined by  $ t=t_{0}$,  yields a nontrivial result. The induced metric is 
\begin{eqnarray}
 &&ds_\Sigma^2=\frac{\ell^2}{\left(\nu^2+3\right) (r-p) (r-q)}dr^{2} 
 +\frac{1}{4} \ell^2 r f d\phi^{2} 
\end{eqnarray}
 in the induced coordinates $ (r,\phi)$.

Now, we can calculate the relevant invariants as given by (\ref{gerekli}) and/or (\ref{gerekli2}) as both invariants work as in the case of the BTZ metric. The explicit expressions are cumbersome, hence we do not depict here; but note that, they both vanish on the event horizon, i.e. 
\begin{eqnarray}
&&{}^{\Sigma}I_{5}(r=r_+)=0,\\
&&{}^{\Sigma}I_{3}(r=r_+)=0.
\end{eqnarray} 
\section{Two examples in 2+1 dimensions }

In 2+1 dimensions, there is a purely quadratic gravity theory (the so called $K$-gravity) defined by the action \cite{Deser}
\begin{equation}
S = \int d^3 x \sqrt{-g} \left ( R_{\mu \nu}R^{\mu \nu}- \frac{3}{8} R^2 \right )
\end{equation}
that admits asymptotically flat and rotating black holes \cite{Barnich,Alkac}. Let us apply the above procedure to detect the event horizons of these black holes. 

\subsection{Asymptotically flat solution of $K$-gravity}

The metric is given as 
\begin{eqnarray}
&&ds^2= -\left(b r-\mu\right)dt^2+\dfrac{1}{b r-\mu}dr^2
+r^2 d\phi^2.
\end{eqnarray}
The usual method shows that there is an event horizon at $r= \mu/b $, as is clear from the metric. This spacetime has two symmetries and has cohomogeneity one which admits two possible hypersurfaces. But one of these, that is reducing along the $\phi$ coordinate, yields a flat space and hence, it is not a viable option. So, we choose the hypersurface defined by $ t= t_0 $. Then the relevant invariants (\ref{gerekli}) and (\ref{gerekli2}) read as 
\begin{equation}
{}^{\Sigma}I_{5}=  \frac{4 b^4 (b r-\mu)}{r^6},
\end{equation}
and
\begin{equation}
{}^{\Sigma}I_{3}=  \frac{b^2 (b r-\mu)}{r^4},
\end{equation}
\vspace{0.4cm}
which vanish on the horizon.

\subsection{Rotating solution of $K$-gravity}

The following metric  which describes a rotating black hole also solves the $K$-gravity \cite{Barnich}
\begin{eqnarray}
&& ds^2=(\mu-br)du^{2} \nonumber\\
&&-2\sqrt{\frac{\left(a^2 b+8 r\right)^2}{a^4 b^2+16 a^2 \mu+64 r^2}}dudr\nonumber\\
&&+a (\mu-b r)dud\phi\nonumber\\
&&+ \left(\frac{a^4 b^2}{64}+\frac{a^2 \mu}{4}+r^2\right) d\phi^{2}.
\end{eqnarray}
The cohomogeneity of this spacetime is one. Therefore, we can detect the event horizon by using the curvature invariants $ {}^{\Sigma}I_{5}$ (\ref{gerekli}) and/or ,$ {}^{\Sigma}I_{3}$ (\ref{gerekli2}). We can induce the spacetime on the constant $ \phi_{0} $ hypersurface  which has 
\begin{eqnarray}
&&{}^{\Sigma}I_{5}=\frac{16777216 a^8 b^4 (b r-\mu) \left(a^2 b^2-8 b r+16 \mu\right)^2}{\left(a^2 b+8 r\right)^{16}}\nonumber\\
&& \times\left(a^2 b^2-4 b r+12 \mu\right)^2 \left(a^4 b^2+16 a^2 \mu+64 r^2\right)
\end{eqnarray}
and
\begin{eqnarray}
&&{}^{\Sigma}I_{3}=\frac{65536 a^4 b^2 (b r-\mu) \left(a^2 b^2-4 b r+12 \mu\right)^2 }{\left(a^2 b+8 r\right)^{10}}\nonumber\\
&&\times \left(a^4 b^2+16 a^2 \mu+64 r^2\right).
\end{eqnarray}
Both invariants vanish, for real values of $r$, on the event horizon $ r = \mu/b$ and change sign there. They also vanish for some other values of $r$ but at these points there is no change of sign.

\section{Kerr Black Hole}

Let us apply the above reasoning to the four dimensional Kerr black hole \cite{Kerr}, which has two Killing symmetries. Therefore, the cohomogeneity of this spacetime is two. By choosing  some special hypersurfaces, we can reduce the cohomogeneity to one. 

The components of the Kerr metric in the Boyer-Lindquist coordinates $(t,r,X,\phi)$ is 
\begin{eqnarray}
   &  &ds^2=- (1-\dfrac{2mr}{\rho^{2}})dt^{2}-\dfrac{4amr(1-X^{2})}{\rho^{2}}dtd\phi\nonumber\\
    &  &+(r^{2}+a^{2}+\dfrac{2a^{2}mr(1-X^{2})}{\rho^{2}})(1-X^{2})d\phi^{2}+\dfrac{\rho^{2}}{\Sigma}dr^{2}\nonumber\\
    &  &+\dfrac{\rho^{2}}{1-X^{2}}dX^{2},
 \end{eqnarray}
  where $ \rho^{2}=r^{2}+a^{2}X^{2} $, $ \Sigma=r^{2}-2mr+a^{2} $, $ X=\cos\theta $, and $ m $ and $ a $ correspond to mass and the rotation parameter, respectively. The hypersurface that will be embedded into the spacetime is defined by,
\begin{equation}
 p\in M,  \qquad    p\in \Sigma \iff X(p)=X_{0}=\text{constant}.
 \end{equation}
The induced metric on the hypersurface can be found by the pulling it back from the spacetime metric. The components of it can be written with respect to the induced coordinates $ (t,r,\phi) $ as
\begin{eqnarray}
  &  &ds_\Sigma^2=- (1-\dfrac{2mr}{\rho^{2}})dt^{2}-\dfrac{4amr(1-X_{0}^{2})}{\rho^{2}}dtd\phi\\
   &  &+(r^{2}+a^{2}+\dfrac{2a^{2}mr(1-X_{0}^{2})}{\rho^{2}})(1-X_{0}^{2})d\phi^{2}
    +\dfrac{\rho^{2}}{\Sigma}dr^{2},\nonumber
 \end{eqnarray}\\
The Kretschmann scalar 
\begin{equation}
{}^{\Sigma} I_1={}^{\Sigma} R_{i j k l} {}^{\Sigma}R^{i j k l}.
\end{equation}\\
of the hypersurface can be found, from which  one can compute 
\noindent
\begin{eqnarray}
{}^{\Sigma} I_{5}=&&\nabla_{\mu}{}^{\Sigma} I_1 \nabla^{\mu}{}^{\Sigma} I_1,
\end{eqnarray}
which for $X_0=0$ yields
\begin{eqnarray}
&&{}^{\Sigma} I_{5}=\frac{20736 M^4 \left(a^2+r (r-2 M)\right)}{r^{16}}
\end{eqnarray}\\
that vanishes on the event horizon given as $r_+= m + \sqrt{ m^2- a^2}$, which matches the result of the coordinate dependent method.

Let us stress that instead of the hypersurface construction given above, if we had used the full spacetime invariant given as 
 \begin{equation}
I_5=\nabla_{\mu}I_{1}\nabla^{\mu} I_{1} ,
\end{equation}
we could not have detected the event horizon with this invariant alone. 
Therefore, in this approach, we have a simplified way to detect the horizon. Another advantage of this approach is that, when it is calculated in the $ X(p)=X_{0}=0 $ hypersurface, the invariant vanishes {\it only} on the event horizon of the black hole and there are no other roots at finite $r$. In the method of \cite{Page}, the norm of the wedge square have other roots in addition to the root indicating the horizon. \\

\section{4+1 MYERS-PERRY BLACK HOLE}
 
Myers and Perry \cite{Myers} found the rotating, massive, Ricci flat black holes in generic $D\ge 5$ dimensions. Here, as an example, we consider the five dimensional solution that can be studied with our method with the help of a single curvature invariant.   In this spacetime, the coordinates are chosen as ($ t,r,X,\phi_{1},\phi_{2} $) where $ X=\cos \theta $. The metric is then, 
\begin{eqnarray}
& & ds^2=-dt^{2}+\dfrac{m}{\Sigma}\left(dt+k(1-X^{2})d\phi_{1}+lX^{2}d\phi_{2}\right)^{2}\nonumber\\
&  &+\dfrac{r^{2}\Sigma}{\Pi-mr^{2}}dr^{2}+\dfrac{\Sigma}{1-X^{2}}dX^{2}+(r^{2}+k^{2})(1-X^{2})d\phi_{1}^{2}\nonumber\\
&  &+(r^{2}+l^{2})X^2d\phi_{2}^{2}
\end{eqnarray}
where
\begin{eqnarray}
&& \Sigma=r^2+k^2X^{2}+l^{2} (1-X^{2}) \nonumber \\
&& \Pi=(r^{2}+k^{2})(r^{2}+l^{2}),
\end{eqnarray}
Here $k,l$ are angular momentum parameters as there are two possible independent angular momenta in five dimensions.
In this coordinate system, the coordinates $ t, \phi_{1}, \phi_{2} $ have associated symmetries. Hence, the cohomogeneity of this spacetime is two. We can induce the spacetime onto the hypersurface of constant $ X $, let say $ X_{0} $.  The induced metric is 
\begin{eqnarray}
& & ds_\Sigma^2=-dt^{2}+\dfrac{m}{\Sigma}(dt+k(1-X_{0}^{2})d\phi_{1}+lX_{0}^{2}d\phi_{2})^{2}\nonumber\\
&  &+\dfrac{r^{2}\Sigma}{\Pi-mr^{2}}dr^{2}+(r^{2}+k^{2})(1-X_{0}^{2})d\phi_{1}^{2}\nonumber\\
&  &+(r^{2}+l^{2})X_{0}^2d\phi_{2}^{2}
\end{eqnarray}
in the induced coordinates $ (t,r,\phi_{1},\phi_{2}) $. Again, one computes the Kretschmann invariant of this hypersurface from which one computes (\ref{gerekli}). The explicit forms of the expressions are complicated, but one can show that ${}^{\Sigma}  I_{5}=0$ at 
\begin{equation}
r_{+}=\frac{\sqrt{\sqrt{\left(k^2+l^2-m\right)^2-4 k^2 l^2}-k^2-l^2+m}}{\sqrt{2}}.
\end{equation}
which is the location of the event horizon.

\section{Conclusions}
Inspired by the recent progress \cite{Lake} in defining the event horizon of black holes  with curvature invariants in $D = 4$ dimensions and $D \ge 4$ dimensions \cite{Page}, we have presented a method which also works for $D=3$ spacetime dimensions. The method of Page and Shoom makes use of the invariants of full spacetime that works well unless the invariants are all constant. Our method adapts their idea, but instead of using the invariants of the full spacetime, we use the invariants of well-chosen hypersurfaces. We have given several examples in three, four and five dimensions where we need only one invariant to detect the black hole horizon. 
In three dimensional black holes, the hypersurface version of the invariant, suggested in \cite{Karlhede} originally for the Schwarzschild black hole, works. Besides this invariant, we provided another one constructed from the gradients of the induced Kretschmann invariant that also works in four and five dimensions.
 Beyond five dimensions, just like the method of \cite{Page}, generically with our method, we need more than one horizon detecting invariant because usually the cohomogeneity is more than two.
For more on the construction of horizon detecting invariants see \cite{Cartan} where Cartan invariants are suggested and see \cite{Gregoris} where they were employed for lower dimensional nonvacuum  black holes such as the charged BTZ metric.


\begin{thebibliography}{10}


\bibitem{Eric} 
  E.~Gourgoulhon and J.~L.~Jaramillo,
  ``A 3+1 perspective on null hypersurfaces and isolated horizons,''
  Phys.\ Rept.\  {\bf 423}, 159 (2006).

	
\bibitem{Lake} M.~Abdelqader and K.~Lake,
``Invariant characterization of the Kerr spacetime: Locating the horizon and measuring the mass and spin of rotating black holes using curvature invariants,''
  Phys.\ Rev.\ D {\bf 91}, no. 8, 084017 (2015).
	
\bibitem{Karlhede} 
  A.~Karlhede, U.~Lindstrom and J.~E.~Aman,
  ``A note on a local effect at the Schwarzschild sphere,''
  Gen.\ Rel.\ Grav.\  {\bf 14}, 569 (1982).
	
\bibitem{Page} 
  D.~N.~Page and A.~A.~Shoom,
  ``Local Invariants Vanishing on Stationary Horizons: A Diagnostic for Locating Black Holes,''
  Phys.\ Rev.\ Lett.\  {\bf 114}, no. 14, 141102 (2015).
	

\bibitem{BTZ} 
  M.~Banados, C.~Teitelboim and J.~Zanelli,
  ``The Black hole in three-dimensional space-time,''
  Phys.\ Rev.\ Lett.\  {\bf 69}, 1849 (1992).

\bibitem{Nutku} 
  Y.~Nutku,
  ``Exact solutions of topologically massive gravity with a cosmological constant,''
  Class.\ Quant.\ Grav.\  {\bf 10}, 2657 (1993).

\bibitem{Gurses} 
  M.~Gurses,
  ``Perfect Fluid Sources in 2+1 Dimensions,''
  Class.\ Quant.\ Grav.\  {\bf 11}, no. 10, 2585 (1994).

\bibitem{Li} 
  D.~Anninos, W.~Li, M.~Padi, W.~Song and A.~Strominger,
  ``Warped AdS(3) Black Holes,''
  JHEP {\bf 0903}, 130 (2009).


\bibitem{DJT} 
  S.~Deser, R.~Jackiw and S.~Templeton,
  ``Topologically Massive Gauge Theories,''
  Annals Phys.\  {\bf 140}, 372 (1982).

\bibitem{Deser}
  S.~Deser,
  ``Ghost-free, finite, fourth order $D=3$ (alas) gravity,''
  Phys.\ Rev.\ Lett.\  {\bf 103}, 101302 (2009).

\bibitem{Barnich} 
  G.~Barnich, C.~Troessaert, D.~Tempo and R.~Troncoso,
  ``Asymptotically locally flat spacetimes and dynamical nonspherically-symmetric black holes in three dimensions,''
  Phys.\ Rev.\ D {\bf 93}, no. 8, 084001 (2016).

\bibitem{Alkac} 
  G.~Alkac, E.~Kilicarslan and B.~Tekin,
  ``Asymptotically flat black holes in 2+1 dimensions,''
  Phys.\ Rev.\ D {\bf 93}, no. 8, 084003 (2016).

\bibitem{Kerr} 
  R.~P.~Kerr,
  ``Gravitational field of a spinning mass as an example of algebraically special metrics,''
  Phys.\ Rev.\ Lett.\  {\bf 11}, 237 (1963).

\bibitem{Myers} 
  R.~C.~Myers and M.~J.~Perry,
  ``Black Holes in Higher Dimensional Space-Times,''
  Annals Phys.\  {\bf 172}, 304 (1986).


\bibitem{Cartan} 
  D.~D.~McNutt, M.~A.~H.~MacCallum, D.~Gregoris, A.~Forget, A.~A.~Coley, P.~C.~Chavy-Waddy and D.~Brooks,
  ``Cartan Invariants and Event Horizon Detection, Extended Version,''
  Gen.\ Rel.\ Grav.\  {\bf 50}, no. 4, 37 (2018).

\bibitem{Gregoris} 
  D.~Gregoris, Y.~C.~Ong and B.~Wang,
  ``Curvature Invariants and Lower Dimensional Black Hole Horizons,''
  Eur.\ Phys.\ J.\ C {\bf 79}, no. 11, 925 (2019).

\end{thebibliography}
\end{document}